## 3.4 The *h*-index

Grischa Fraumann and Rüdiger Mutz

**Abstract:** The *h*-index is a mainstream bibliometric indicator, since it is widely used in academia, research management and research policy. While its advantages have been highlighted, such as its simple calculation, it has also received widespread criticism. The criticism is mainly based on the negative effects it may have on scholars, when the index is used to describe the quality of a scholar. The "h" means "highly-cited" and "high achievement", and should not be confused with the last name of its inventor, Hirsch. Put simply, the *h*-index combines a measure of quantity and impact in a single indicator. Several initiatives try to provide alternatives to the *h*-index to counter some of its shortcomings.

**Keywords:** *h*-index, metrics, author-level metrics, indicators, productivity, publications, citation impact.

### Introduction

The *h*-index was developed by Jorge Hirsch, a physicist at the University of California at San Diego, who published the concept in the *Proceedings of the National Academy of Sciences of the USA* (Hirsch, 2005). The *h*-index was defined by Hirsch as follows: "A scientist has index *h* if *h* of his or her $N_p$ papers have at least *h* citations each and the other ($N_p - h$) papers have $\leq h$ citations each." (Hirsch, 2005) The "h" means "highly-cited" and "high achievement", and should not be confused with the last name of its inventor (Hirsch and Buela-Casal, 2014; Schubert and Schubert, 2019). Put simply, the *h*-index "combines a measure of quantity and impact in a single indicator" (Costas and Bordons, 2007).

While the *h*-index was proposed in 2005 and is considered a classical bibliometric indicator, there is still an ongoing debate on its value in bibliometrics and in the scholarly community in general. Due to its popularity, it has even been called a mainstream bibliometric indicator (Costas and Franssen, 2018). It is one of the most well-known indicators, but it has received negative and positive judgements alike.

The development of indicators was part of a shift from higher level entities (e.g., countries, institutions, journals) towards the bibliometrics of individual researchers

**Grischa Fraumann**, Research Assistant at the TIB Leibniz Information Centre for Science and Technology in the R&D Department, and PhD Fellow at the University of Copenhagen in the Department of Communication. He is also a Research Affiliate at the "CiMetrias: Research Group on Science and Technology Metrics" at the University of São Paulo (USP), grischa.fraumann@tib.eu

**Rüdiger Mutz**, Senior Researcher at Center for Higher Education and Science Studies (CHESS), University of Zurich, ruediger.mutz@uzh.ch

(Costas, van Leeuwen, and Bordons, 2010; Hicks et al., 2015). This shift was related to how single researchers' research outputs should be measured via quantitative indicators. How analyses should be conducted on an individual level was and still remains an open question in bibliometrics (Bornmann and Marx, 2014). The relatively simple *h*-index calculation has contributed to its widespread use (Sugimoto and Larivière, 2018), and it was promoted in several journals and news outlets in the beginning (Ball, 2005). Hirsch's original work from 2005, for example, has been cited 4,530 times according to the bibliographic databases *Scopus* (accessed October 25, 2019), 3,999 times according to *Web of Science* (accessed October 25, 2019) and 5,240 times according to *The Lens* (accessed September 23, 2019). Due to the high number of publications on the *h*-index (Waltman, 2016), this book chapter focuses on some central topics of *h*-index research and refers mostly to comprehensive literature reviews (Costas and Bordons, 2007).

## *h*-index History

The quantitative distribution of scholarly works on the h-index over a period of 13 years may provide an insight into its historical development. Therefore, a bibliometric analysis was conducted based on the bibliographic database *The Lens* (Jefferson et al., 2018) to query all works for "*h*-index" in the title, abstract, keyword or field of study. All publications from January 1, 2005, the year of Hirsch's original h-index publication, to December 31, 2018 were retrieved (accessed September 22, 2019).

The total number of publications was 3,817, and the number increased more or less steadily each year until it dropped slightly in 2009 and once again in 2013 (Figure 1). The numbers from 2014 until 2018 are year by year almost the same, about 400 documents per year.

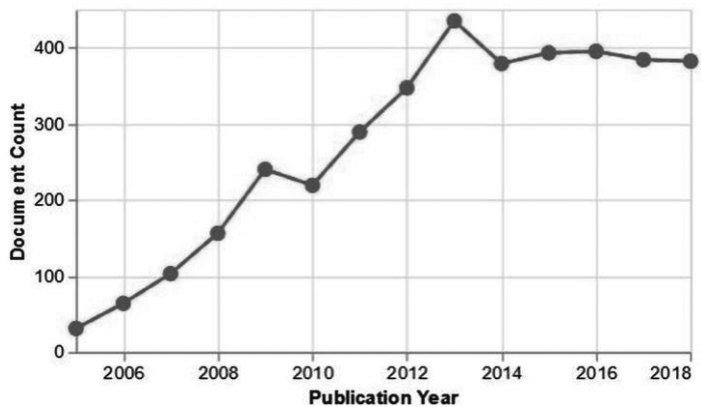

**Figure 1:** Scholarly works over time on the *h*-index (source: *The Lens*).

The *h*-index has been applied in several academic disciplines, such as physics, biomedicine, information science and business studies (Costas and Bordons, 2007). However, a comparison between disciplines by using the *h*-index is not recommended (Hirsch, 2005; van Leeuwen, 2008) because of different citation practices, among others. There are attempts to put the *h*-index in a common scale in order to make inter-field comparisons possible (Iglesias and Pecharromán, 2007).

Due to the ever increasing number of online sources that provide bibliometric data, the possibility to calculate and/or display the *h*-index has also changed dramatically (Costas and Franssen, 2018; Teixeira da Silva and Dobránszki, 2018). For example, the *h*-index is prominently placed on *Google Scholar* profiles (Costas and Wouters, 2012; Sugimoto and Larivière, 2018) and also in the *Web of Science* and *Scopus* (Leydesdorff, Bornmann, and Opthof, 2019).

The scope of application of the *h*-index has been extended, for instance, to journals (e.g., Braun, Glänzel, and Schubert, 2006) and to research groups (e.g., van Raan, 2006). Several variants of the *h*-index have been developed to address specific limitations of the *h*-index (see section 4), for example, the *g*-index as the most prominent one (Egghe, 2006).

## *h*-index Concept

In order to enhance the understanding of the basic *h*-index concept, a graphical derivation of the index could be helpful (Alonso et al., 2009). In Figure 2 the so-called rank frequency distribution for a researcher is shown. His or her publications are sorted in descending order according to their citations. The publication with the highest citation is ranked first, then the publication with the second highest citation and so on. The *h*-index corresponds to intersection between the rank-frequency distribution and the 45° degree line, where the number of papers is equal to the number of publications. For the specific researcher the *h*-index amounts to 22. The *h*-core comprises 22 publications, which contribute to the *h*-index and are cited at least 22 times. Neither the number of citations the highly cited papers receive nor the papers outside the *h*-core are of importance.

Several advantages of the *h*-index were discussed in the literature (Alonso et al., 2009; Bornmann and Daniel, 2005; Bornmann and Daniel, 2007; Costas and Bordons, 2007). It is a robust indicator in the sense that it is rather insensitive to lowly cited papers. It is an objective indicator and might play a certain role together with other indicators and expert judgements during funding or promotion decisions. It claims to perform better than any single indicator. Hirsch himself postulated that the *h*-index has a predictive capacity for researchers' careers to a greater extent than conventional citation indicators (Alonso et al., 2009; Hirsch, 2007). Bornmann and Daniel (2005), for instance, found that the *h*-index for successful applicants for post-doctoral biomedicine fellowships was significantly higher than for non-successful applicants.

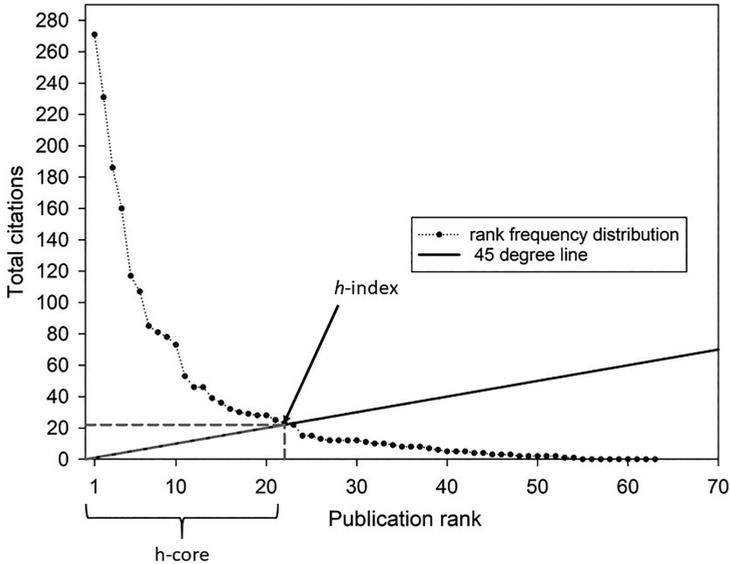

**Figure 2:** Graphical representation of the *h*-index.

## *h*-index Problems and Alternatives

The *h*-index is widely used by researchers, for example in the medical sciences (Cronin and Sugimoto, 2014), but it has also been widely criticised (Bornmann, Mutz, and Daniel, 2010; Hauschke, 2019; Waltman and van Eck, 2012). The *h*-index's widespread availability online has the potential for problematic usage in research evaluation, which should never rely on a single indicator (Costas and Bordons, 2007). Conversely, this simplicity and objectivity is sometimes also seen as an advantage of the *h*-index (Alonso et al., 2009). However, the indicator could also lead to questionable self-citation practices carried out to increase the *h*-index (Costas and Bordons, 2007). For example, the *h*-index may be problematic due to the Matthew effect, whereby a high-profile researcher may attract more and more citations based on her or his displayed *h*-index (Alonso et al., 2009).

Costas and Bordons pointed out the problem of size dependency of the *h*-index. Imagine the following scenario:

> [S]cientist "A" with 10 documents cited 10 times each would have an h-index of 10; whereas scientist "B", with 5 documents which were cited 200 times each, would only achieve an h-index of 5. Scientist "B" publishes fewer documents, but their impact is much higher than the other's (i.e., a higher citation per document rate). Scientist "A" publishes many more documents, albeit with a lower impact. Despite this, according to the h-index, scientist A would be regarded as much more successful than "B" (Costas and Bordons, 2007).

The example above illustrates that researchers with a higher *h*-index might be those who have a high quantity of publications with average citations compared to researchers that publish less but have a few highly-cited publications (Costas and Franssen, 2018).

The *h*-index is also influenced by the length of a researcher's career or lifetime citedness (Alonso et al., 2009; Ball, 2017; Bornmann and Marx, 2011; Costas and Bordons, 2007; Hicks et al., 2015). It combines quantity as well as impact in one indicator or single number (Hirsch, 2005). Colloquially, this concept might also be called an all-in-one metric (Bornmann and Marx, 2011), while its value cannot decrease because it relies on the number of publications and citations (Gingras, 2016). Obviously, this focus on lifetime citedness may be disadvantageous to early career researchers. The merger of two concepts in one indicator has also been criticised (Sugimoto and Larivière, 2018), even if Hirsch's original proposal was to simplify "to [a] great extent the characterization of researchers' scientific output" (Alonso et al., 2009). Furthermore, the *h*-index might be manipulated, for example in *Google Scholar* (Costas and Wouters, 2012; Gingras, 2016). Additionally, it does not distinguish between negative or positive citations (Alonso et al., 2009), although such a distinction is difficult to achieve as it requires natural language processing (Teufel, Siddharthan, and Tidhar, 2006). A researcher could have a different *h*-index depending on the bibliographic database (Bar-Ilan, 2008; Hicks et al., 2015) and academic disciplines (Hicks et al., 2015). Lists of researchers that display their *h*-index[1] may be insightful, but the use of the *h*-index in hiring and promotion decisions might be problematic (Hicks et al., 2015). According to the Leiden Manifesto (Hicks et al., 2015, Principle 7), the San Francsico Declaration on Research Assessment (DORA) (Cagan, 2013, Principle 3), and the Hong Kong Principles (Moher et al., 2020, Principle 1), research assessment of individual researcher should consider a broad range of bibliometric measures, not only a single indicator, such as the *h*-index. In general, quantitative research assessment should support not replace qualitative judgements of experts (e. g., peer review) (Hicks et al., 2015).

Alternative types of indicators have been studied, such as the *g*-index (Costas and Bordons, 2008; Egghe, 2006), the *hg*-index, the *A*-index, and the *m*-index (Alonso et al., 2009). Meta-analyses have been carried out to study correlations between the *h*-index and its variants (Bornmann et al., 2011), such as the *g*-index. Strikingly, a 2011 meta-analysis concluded that 35 out of 37 variants seemed to duplicate the *h*-index, except for the modified impact index (*MII*) and *m*-index (Bornmann et al., 2011). This means that most variants are highly correlated with the *h*-index (Bornmann and Mutz et al., 2009). Bornmann et al. also categorised the variants into two groups, namely the "impact of the productive core" and the "quantity of the productive core". The productive core refers to the most received citations (Bornmann, Mutz, and Daniel, 2008). Studies have also concluded that the *h*-index and its var-iants are not needed compared to standard bibliometric measures, such as number of publications and total citation counts (Bornmann, Mutz, and Daniel, 2009). Due to these findings, some scholars suggested not developing new variants of the *h*-index any longer (Bornmann and Marx et al., 2009) but rather to complement or enrich the *h*-index with additional information (Bornmann, Mutz, and Daniel, 2010). Variants of the *h*-index, such as the ones mentioned above, are also regarded as superficial enhancements (Waltman, 2016). From a bibliometric perspective, the *h*-index seems to have no analytical value as such (Leydesdorff, Bornmann, and Opthof, 2019), despite its frequent use in academia, research management and policy. Initiatives are underway to employ indicators by involving the scholarly community (Hauschke, Cartellieri, and Heller, 2018), to overcome the evaluation gap between what is measured by indicators and what is valued by researchers (Heuritsch, 2018; Wouters, 2017).

---

[1] https://www.webometrics.info/en/hlargerthan100 (July 15, 2020).

## Conclusions

This article has provided an overview of the *h*-index, described its applications, reviewed several studies that scrutinise the *h*-index and discussed its advantages and disadvantages. Developed by Hirsch in 2005 as an "index to quantify an individual's scientific output" (Hirsch, 2005), the *h*-index is still being debated almost 15 years later. A vast amount of literature on the *h*-index is available. As such, it is perhaps one of the most studied topics in bibliometrics and scientometrics, and it has had an influence on the scholarly community as a whole and even beyond on research management and policy. Several variants of the *h*-index have been developed over the years, but a significant improvement would be to provide additional information and indicators and explain the *h*-index context. The *h*-index certainly also has some advantages, such as its simple calculation and wide availability. A problem that remains is its isolated use in research evaluation, potential comparisons among academic disciplines and the fact that it is taken by some at face value to judge the quality of a researcher's work. Finally, the display of the *h*-index in bibliographic databases, such as *Google Scholar*, codifies this indicator (Sugimoto and Larivière, 2018) and normalises its perception. The future of the *h*-index is still being debated, but the negative assessments seem to be in the majority.

**Acknowledgements**
This chapter was funded by the German Federal Ministry of Education and Research (BMBF) under grant numbers 01PU17019 (ROSI – Reference Implementation for Open Scientometric Indicators) and 16PGF0289 (BMBF Post-Grant-Fund).